\documentclass[conference]{IEEEtran}
\IEEEoverridecommandlockouts
\usepackage{cite}
\usepackage{lscape} 
\usepackage{longtable}
\usepackage{booktabs} 
\usepackage{multirow} 
\usepackage{amsmath,amssymb,amsfonts}
\usepackage{algorithmic}
\usepackage{graphicx}
\usepackage{textcomp}
\usepackage{subcaption}
\usepackage{ragged2e,array,booktabs}
\usepackage{siunitx}
\usepackage{array, ltablex, multirow}
\usepackage{tabularx}
\usepackage{colortbl}
\usepackage{hhline}
\usepackage{acronym}
\usepackage[latin9]{inputenc}
\usepackage[table]{xcolor}
\usepackage{collcell}
\usepackage{hhline}
\usepackage{pgf}
\usepackage{multirow}
\usepackage{soul}

\def\colorModel{hsb} 

\newcommand\ColCell[1]{
  \pgfmathparse{#1<50?1:0}  
    \ifnum\pgfmathresult=0\relax\color{white}\fi
  \pgfmathsetmacro\compA{0}      
  \pgfmathsetmacro\compB{#1/100} 
  \pgfmathsetmacro\compC{1}      
  \edef\x{\noexpand\centering\noexpand\cellcolor[\colorModel]{\compA,\compB,\compC}}\x #1
  } 
\newcolumntype{E}{>{\collectcell\ColCell}m{0.4cm}<{\endcollectcell}}  

\newcommand{\fakepar}[1]{\vspace{1mm}\noindent\textbf{#1.}}

\def\BibTeX{{\rm B\kern-.05em{\sc i\kern-.025em b}\kern-.08em
    T\kern-.1667em\lower.7ex\hbox{E}\kern-.125emX}}

\acrodef{adc}[ADC]{Analog-to-Digital Converter}
\acrodef{tmd}[TMD]{Transport Mode Detection}
\acrodef{keh}[KEH]{Kinetic Energy Harvesting}
\acrodef{iot}[IoT]{Internet of Things}
\acrodef{rf}[RF]{Random Forest}
\acrodef{dt}[DT]{Decision Tree}
\acrodef{svm}[SVM]{Support Vector Machine}
\acrodef{knn}[KNN]{K-Nearest Neighbor}
\acrodef{nb}[NB]{Naive Bayes}
\acrodef{seh}[SEH]{Solar Energy Harvesting}
\acrodef{apr}[APR]{Acquisition Power Ratio}
\acrodef{led}[LED]{Light Emitting Diode}
\acrodef{cv}[CV]{Cross Validation}
\acrodef{rfe}[RFE]{Recursive Feature Elimination}
\acrodef{smote}[SMOTE]{Synthetic Minority Over-sampling Technique}

\begin{document}
\title{Towards Energy Positive Sensing using\\Kinetic Energy Harvesters \\
}

\author{\IEEEauthorblockN{Muhammad Moid Sandhu\textsuperscript{1,2}, Kai Geissdoerfer\textsuperscript{3}, Sara Khalifa\textsuperscript{2,4}, Raja Jurdak\textsuperscript{5,2}, Marius Portmann\textsuperscript{1}, Brano Kusy\textsuperscript{2}}
\IEEEauthorblockA{\textsuperscript{1}School of Information Technology \& Electrical Engineering, The University of Queensland, Brisbane, Australia \\
\textsuperscript{2}Data61, Commonwealth Scientific and Industrial Research Organization (CSIRO), Australia\\
\textsuperscript{3}Networked Embedded Systems Lab, TU Dresden, Germany\\
\textsuperscript{4}School of Computer Science \& Engineering, University of New South Wales, Sydney, Australia \\
\textsuperscript{5}School of Electrical Engineering \& Computer Science, Queensland University of Technology, Brisbane, Australia \\
{\small m.sandhu@uqconnect.edu.au, kai.geissdoerfer@tu-dresden.de, sara.khalifa@data61.csiro.au, r.jurdak@qut.edu.au,}\\ {\small marius@itee.uq.edu.au, brano.kusy@csiro.au}}
}

\maketitle

\begin{abstract}
Conventional systems for motion context detection rely on batteries to provide the energy required for sampling a motion sensor.
Batteries, however, have limited capacity and, once depleted, have to be replaced or recharged.
\ac{keh} allows to convert ambient motion and vibration into usable electricity and can enable batteryless, maintenance free operation of motion sensors.
The signal from a \ac{keh} transducer correlates with the underlying motion and may thus directly be used for context detection, saving space, cost and energy by omitting the accelerometer.
Previous work uses the open circuit or the capacitor voltage for sensing without using the harvested energy to power a load.
In this paper, we propose to use other sensing points in the \ac{keh} circuit that offer information-rich sensing signals while the energy from the harvester is used to power a load.
We systematically analyze multiple sensing signals available in different \ac{keh} architectures and compare their performance in a transport mode detection case study.
To this end, we develop four hardware prototypes, conduct an extensive measurement campaign and use the data to train and evaluate different classifiers.
We show that sensing the harvesting current signal from a transducer can be energy positive, delivering up to ten times as much power as it consumes for signal acquisition, while offering comparable detection accuracy to the accelerometer signal for most of the considered transport modes.
\end{abstract}
\begin{IEEEkeywords}
Kinetic, Energy harvester, Wearables, IoT, Context detection, Energy positive sensing, Sensing points, Signals
\end{IEEEkeywords}
\section{Introduction}
Recent technological advances have made pervasive wearable devices for human health~\cite{cheol2018wearable}, fitness~\cite{scalise2018wearables}, and activity~\cite{hegde2017automatic} monitoring possible. According to~\cite{dias2018wearable}, the wearable devices market is currently having a worldwide revenue
of around \$26 billion, and is expected to reach almost \$73 billion in 2022.
Monitoring an individual's daily activities enables location-based services~\cite{castro2013taxi}, travel route planning~\cite{liu2014exploiting}, surveillance~\cite{kumari2017increasing}, and human computer interaction~\cite{welch2018wearable}.

\begin{figure}[t!]
\centering
\includegraphics[width=8.5cm, height=2cm]{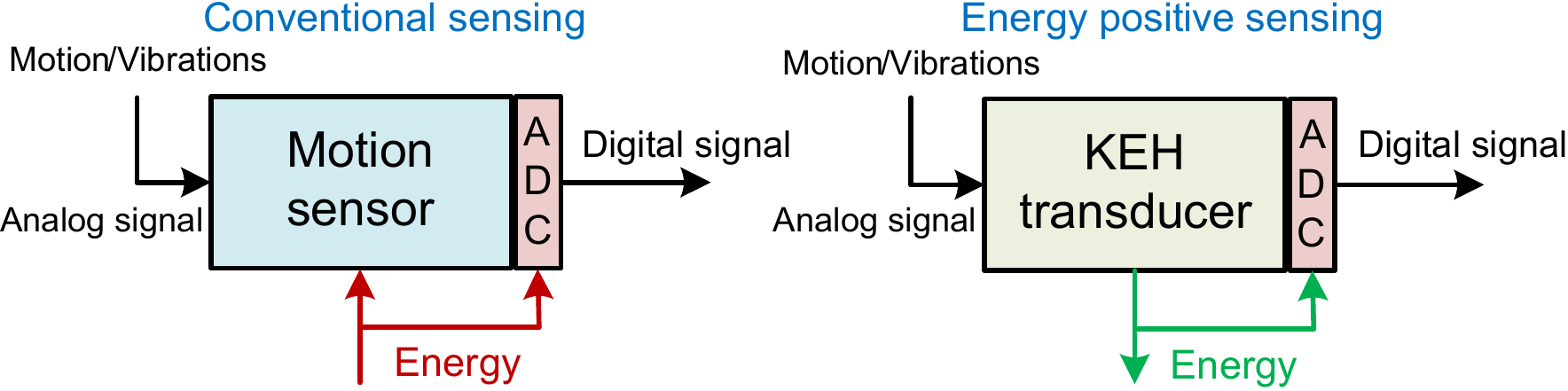}\vspace{-0.2cm}
\caption{Conventional (energy negative) vs energy positive sensing}
\label{fig:energy_positive_negative}
\vspace{-0.65cm}
\end{figure}

\fakepar{Motivation}
Conventional motion sensors such as pedometers, accelerometers, inclinometers, and gyroscopes~\cite{taraldsen2012physical} consume energy for converting the physical phenomenon into an analog signal as well as for converting that analog signal into its digital form, using an \ac{adc} as portrayed in Fig.~\ref{fig:energy_positive_negative}.
Naturally, activity sensors for motion context detection must be portable, which poses a challenge for their power supply.
Batteries are bulky, expensive and need to be either replaced or recharged, both of which require access to the device and human intervention.

Kinetic Energy Harvesters (KEHs) convert ambient vibration, stress or motion into usable electrical energy using piezoelectric, electrostatic or electromagnetic transducers.
They can replace batteries, realizing a truly pervasive, sustainable \ac{iot}, consisting of trillions of tiny and economical sensors/devices without generating tons of toxic waste~\cite{hester2017future}.
The physical effects that are exploited for \ac{keh} are the same as used in sensors that measure dynamic mechanical variables, like acceleration~\cite{khalifa2017harke}.
Using \ac{keh} transducer simultaneously as a sensor and as an energy source can replace the conventional accelerometer to reduce system complexity, cost and energy consumption.
If the harvested energy is higher than the energy required for sampling the signal with an \ac{adc}, the additional energy may even be used to power other parts of the system, and the sensor is \textit{energy positive}.

\fakepar{State-of-the-art}
Previous work falls short of solving the key challenges towards \textit{energy positive sensing}.
Most studies~\cite{khalifa2015energy, khalifa2017harke, 8730510, umetsu2019ehaas} explore the sensing potential of the \ac{keh} open circuit voltage, i.e., without actually harvesting energy.
Lan et al.~\cite{lan2017capsense} employ the charging rate of a capacitor connected to a \ac{keh} transducer inside a shoe sole as a sensing signal for human activity recognition.
Their system, however, uses two separate transducers and capacitors for sensing and energy harvesting.
Although it has been shown that sampling the open circuit voltage of a \ac{keh} transducer can save energy compared to using an accelerometer, a second transducer significantly adds to the weight and cost of the system.
The full potential of \ac{keh}-based sensing is only unleashed, when the same transducer is used simultaneously as power supply and sensor.

Ma et al.~\cite{ma2018sehs} make a first step towards simultaneous sensing and energy harvesting by sampling the transducer voltage while storing harvested energy in a capacitor.
They describe the resulting \textit{interference problem}, where the transducer voltage is enveloped in the capacitor voltage, affecting the quality of the sensing signal.
They propose a filter to mitigate this effect, however, their proposed filter only considers the charging curve of the capacitor, neglecting the effects of a dynamic load consuming energy from the capacitor.
It is thus not applicable in practical scenarios, where the harvested energy is used to power the system.

\fakepar{Contribution}
In this paper, we present a system architecture for \textit{energy positive sensing}, where a single transducer is used for sensing while simultaneously powering a dynamic load.
We systematically explore various sensing signals in converter-less and converter-based energy harvesting circuits in order to find a high quality sensing signal that is less affected by the interference problem. To evaluate the end-to-end performance of the proposed system, we collect extensive \ac{keh} data from various transport modes, including ferry, train, bus, car, tricycle and pedestrian movement.
We extract the dominant feature set, implement five classifiers and compare the \ac{tmd} accuracy between the sensing signals.
We find that sensing the harvesting current signal in the converter-based design can be energy positive, delivering up to ten times as much power as it consumes for signal acquisition, while offering comparable detection accuracy to the accelerometer signal for most of the considered transport modes.

The key contributions of this paper are summarized below:
\begin{itemize}
    \item We present the first complete architecture for sampling the signal from a \ac{keh} transducer, while simultaneously powering a dynamic load from the harvested energy.
    \item We systematically explore multiple sensing signals, including current and voltage in two different energy harvesting circuit designs.
    \item We compare the classification performance between various \ac{keh} signals and a 3-axis accelerometer signal in a \ac{tmd} case study. We find that although the interference problem affects the pattern of the harvesting voltage signal, the achievable accuracy in motion context detection is not highly affected.
    \item We show that the harvesting current signal outperforms all other KEH signals and achieves classification performance on par with the accelerometer signal with two-fold lower energy consumption, while generating up to 10 times as much power as it consumes for signal acquisition.
\end{itemize}
The remainder of the paper is organized as follows. Sections~\ref{System_Architecture},~\ref{Hardware_Development} and~\ref{Transport_Mode_Detection_Algorithm} present system architecture, system design and \ac{tmd} as a case study respectively. Section~\ref{results} describes the achieved results and Section~\ref{Literature_Review} discusses the state-of-the-art. Finally, Section~\ref{Conclusion_and_Future_Work} concludes the paper.

\section{System Architecture}
\label{System_Architecture}
\begin{figure}[t!]
\centering
\includegraphics[width=9cm, height=2.1cm]{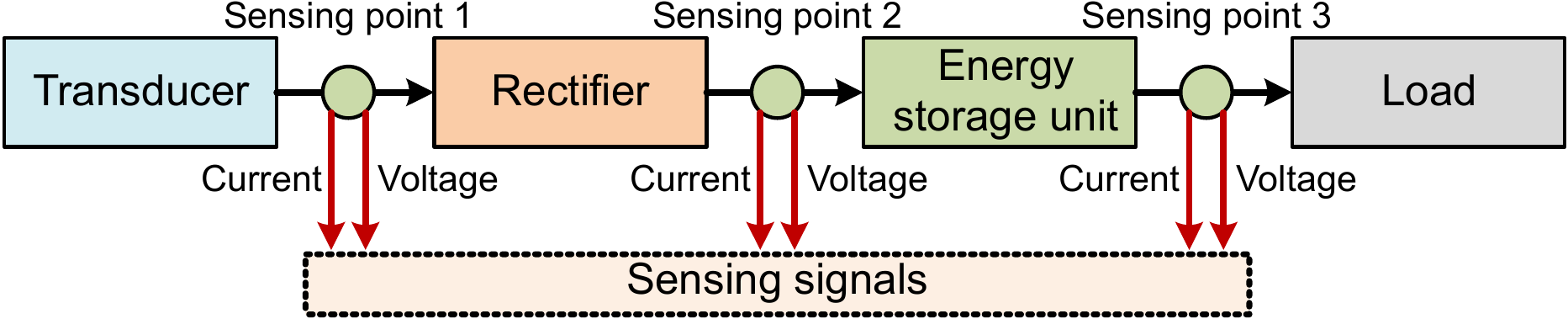}
\caption{Architecture of simultaneous sensing and energy harvesting with the availability of multiple sensing points}
\label{sensing_points}
\vspace{-0.65cm}
\end{figure}

The principal building blocks of the proposed system architecture are shown in Fig.~\ref{sensing_points}.
A transducer converts the kinetic energy into electrical energy and a rectifier is used to rectify the harvesting AC voltage.
The rectified voltage charges a capacitor either directly or through a DC-DC converter.
The energy from the capacitor is used to power a load, for example, the signal acquisition circuit, microcontroller, or a transceiver.
We detail the main characteristics of the system architecture in the following subsections.
\subsection{Simultaneous sensing and energy harvesting}
Due to high source impedance~\cite{kalantarian2015monitoring}, the voltage across a transducer changes dramatically when current flows, i.e., when closing the circuit to extract energy.
When connecting a capacitor to the output of the rectifier, the capacitor voltage envelops the harvesting voltage and changes its pattern compared to the open circuit configuration.
This has been described as the interference problem and previous work proposes a filtering algorithm based on the capacitor voltage to reduce the effect on sensing signal quality~\cite{ma2018sehs}.

Our architecture includes not only a capacitor, but also an intermittently powered load that uses the energy stored on the capacitor, as shown in Fig.~\ref{sensing_points}.
The load reflects the behaviour of a typical batteryless sensor node that switches on when enough charge has been accumulated in a small capacitor to, for example, sample, store, process or transmit the \ac{keh} data.
The load discharges the capacitor and its dynamic behaviour thus also distorts the harvesting voltage waveform, as discussed in Section~\ref{impact_of_energy_harvesting}.
In contrast to previous work~\cite{ma2018sehs}, which uses custom filters to mitigate the effect of the capacitor on the harvesting AC voltage, we explore the potential of other sensing signals, which do not suffer from the interference problem, allowing us to 1) reduce cost in terms of energy, delay and computational complexity by omitting these additional filter stages; and 2) exclude the hard-to-predict effects of dynamic load behaviour on sensing signal quality.
\subsection{Energy positive sensing}

In order to highlight the difference between conventional sensing and \ac{keh}-based sensing, we categorize sensing devices into two classes based on their energy profile; energy negative sensors and \textit{energy positive sensors}.
The key difference between a conventional motion sensor and a \ac{keh} transducer is that the former consumes energy to convert the kinetic energy to an analog signal, whereas the latter generates energy, as shown in Fig.~\ref{fig:energy_positive_negative}.
Both types of sensors require an \ac{adc} to convert the analog signal to its digital form.
Conventional motion sensors (such as accelerometers) are thus always \textit{energy negative} and need to replenish energy regularly for their uninterrupted operation.
\ac{keh}-based sensors, on the other hand, can be energy negative or \textit{energy positive} depending on the amount of harvested energy relative to the energy consumed for acquiring the \ac{keh} signal.
If the harvested energy is higher than the energy required for signal acquisition, it is called \textit{energy positive sensing}.

\begin{figure}[t!]
\centering
\includegraphics[width=9cm, height=3.5cm]{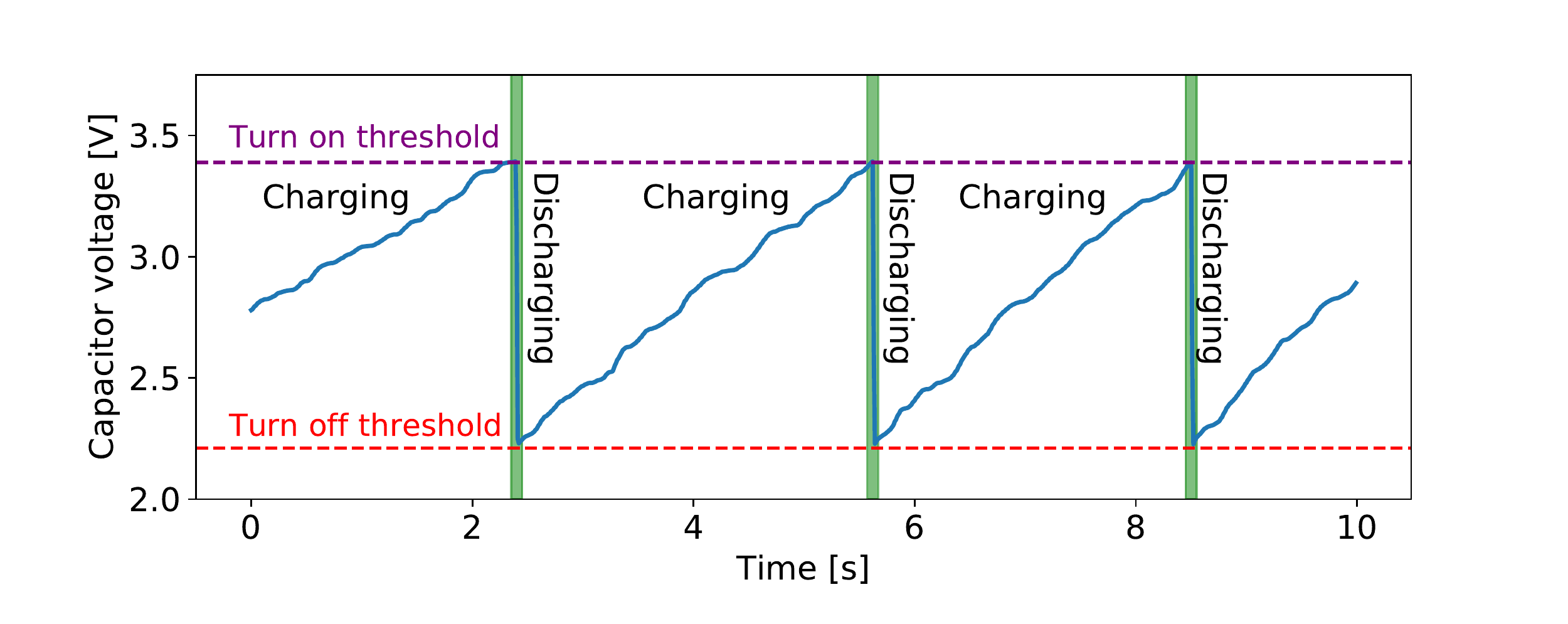}\vspace{-0.3cm}
\caption{Capacitor voltage in intermittently powered sensors}
\label{fig:transiently_powered_sensors}
\vspace{-0.65cm}
\end{figure}
\subsection{Exploring multiple sensing points}
Previous works employ the open circuit AC voltage~\cite{khalifa2017harke, umetsu2019ehaas, 8730510} from the energy harvester or capacitor voltage~\cite{lan2017capsense} for extracting context information. There are various sensing points in the energy harvesting circuit that offer two types of sensing signals i.e., voltage and current which contain context information. Sensing points 1, 2 and 3 in Fig.~\ref{sensing_points} capture the current and voltage signals at the transducer, rectifier and energy storage unit, respectively. We evaluate various \ac{keh} signals by comparing the information content between them, using different designs of the energy harvesting circuit.
\section{System Design}
\label{Hardware_Development}
In this section, we discuss design options for simultaneous sensing and energy harvesting, present our hardware prototypes and analyze the \ac{keh} signals.
\subsection{Hardware designs for \ac{keh} sensing and energy harvesting}

We employ a piezoelectric transducer to convert ambient kinetic energy into electrical energy.
Under mechanical stress, it generates an electric field with alternating polarity~\cite{khalifa2017harke}.
Most previous works on \ac{keh} sensing~\cite{khalifa2015energy, khalifa2017harke, 8730510, umetsu2019ehaas} directly use this open circuit AC voltage as a sensing signal.
To extract usable electrical energy from the transducer, the circuit has to be closed, such that current can flow.
The harvesting current is typically alternating and thus needs to be rectified by, for example, a full bridge rectifier, consisting of four diodes.

\fakepar{Batteryless design}
\label{sec:batteryless_design}
Traditional energy harvesting systems use large rechargeable batteries in order to compensate for variations in the harvested energy~\cite{geissdoerfer2019preact}.
Batteryless devices instead use only a tiny capacitor to accumulate just enough energy to support the largest atomic operation~\cite{gomez2016dynamic}, for example, transmitting a packet over a wireless link.
This works well when the temporal application requirements are well aligned with temporal energy availability, as in \textit{energy positive sensors}.
The sensor only provides useful data, when it also provides energy, allowing to dramatically reduce system cost and size by omitting large batteries.
As the amount of harvested energy is usually too small to support perpetual operation of the device, it resorts to what is known as intermittent execution (shown in Fig.~\ref{fig:transiently_powered_sensors}):
The device remains completely off, while the capacitor accumulates charge from the harvester.
Once the $V_{thr}^{on}$ threshold is reached, the device switches on and executes, quickly draining the capacitor until the $V_{thr}^{off}$ threshold is reached and the device is powered off again.
In contrast to previous work, we discuss the effect of this dynamic load behaviour in the context of the interference problem in Section~\ref{impact_of_energy_harvesting}.

There are two main design options for charging the capacitor which are described in detail below:
\begin{figure}[t!]
\centering
\includegraphics[width=8.5cm, height=6cm]{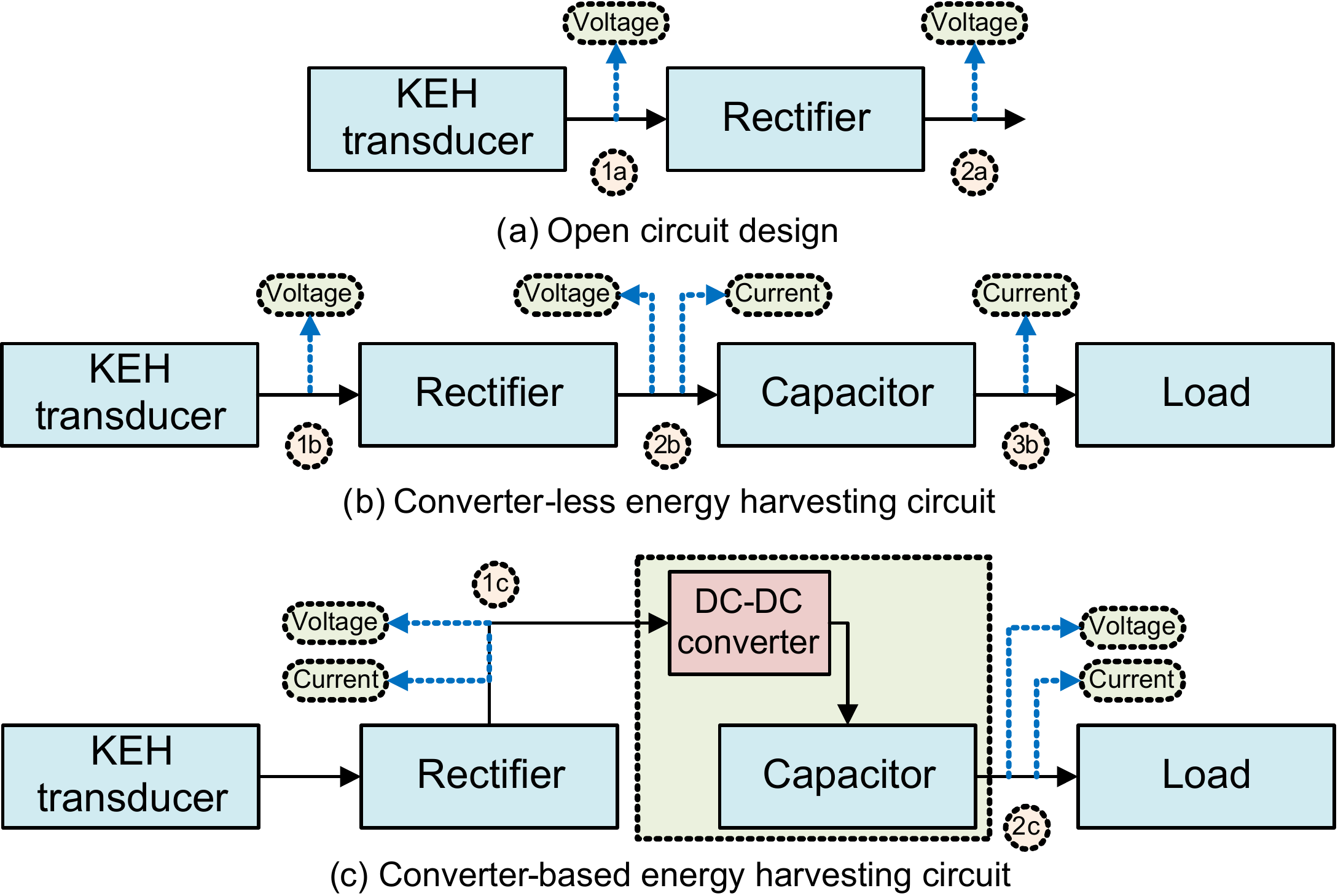}
\caption{Illustration of hardware prototypes developed for \ac{keh} data collection (a) Open circuit design that collects AC and rectified voltages, and (b) Converter-less and (c) Converter-based designs that sample the current and voltage signals at various sensing points in the circuit}
\label{sensing_harware}
\vspace{-0.65cm}
\end{figure}

\fakepar{Converter-less design}
In a converter-less design, the capacitor is placed in parallel to the rectifier.
Therefore, the voltage across the transducer is given by:

\begin{equation}
    v_{AC} = v_{cap}+2\cdot v_d
    \label{eq:envel_formation}
\end{equation}
where, $v_{cap}$ is the capacitor voltage and $v_d$ is the voltage drop across the corresponding diode in the full wave rectifier.
If the open circuit voltage is higher than the voltage on the capacitor, $v_d$ is approximately constant (one diode drop, typically \SI{700}{\milli\volt}) and thus, the voltage across the transducer equals the capacitor voltage plus two diode drops.
This may result in low energy yield.
For example, if the capacitor voltage is \SI{3}{\volt} and the input vibration is low, then the open circuit voltage of the transducer may be less than \SI{3}{\volt}.
In this case, no current can flow and thus energy that could have potentially been harvested is wasted.
The dependence between capacitor voltage and transducer voltage has important implications for the sensing signal quality that will be discussed in Section~\ref{impact_of_energy_harvesting} in detail.

\fakepar{Converter-based design}
In a converter-based design, a DC-DC boost-converter is placed between the rectifier and the capacitor.
This allows to optimize the operating point (i.e., harvesting voltage) of the transducer independent of the voltage on the capacitor.
For example, the converter can be configured to regulate the voltage at its input to \SI{100}{\milli\volt} by dynamically controlling the current flow from the transducer.
This allows to extract energy for charging the capacitor from the transducer even under very low motion or vibrations.
The decoupling of the transducer from the capacitor and load has another important effect: while the harvesting voltage is kept constant by the regulator and thus does not contain context information, the current changes approximately linearly with the kinetic energy input, yielding a high quality sensing signal that is not affected by the interference problem~\cite{ma2018sehs}.
In this document, we refer to DC-DC boost converters, DC-DC converters and converters interchangeably.
\begin{figure}[t!]
\centering
\includegraphics[width=9cm, height=2.5cm]{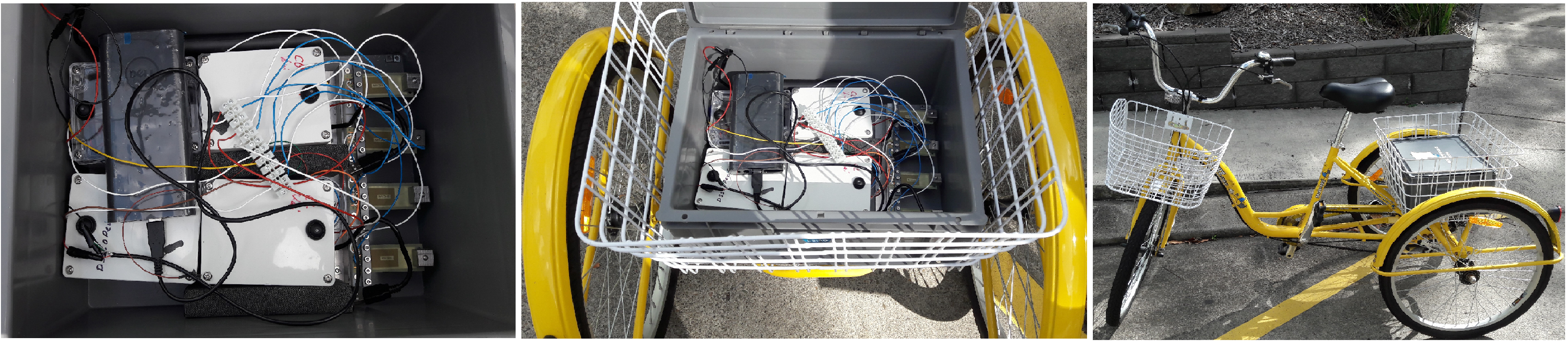}\vspace{-0.1cm}
\caption{Experimental setup for data collection using tricycle}
\label{fig:experiment_setup_bike}
\vspace{-0.45cm}
\end{figure}
\subsection{Prototyping and experimental setup}
For our study, we design four different \ac{keh} hardware prototypes for collecting data from three types of \ac{keh} circuits: Open circuit (Fig.~\ref{sensing_harware}a), converter-less (Fig.~\ref{sensing_harware}b) and converter-based (Fig.~\ref{sensing_harware}c). For the converter-less design, we use two prototypes; one measures the voltage signals, the other measures the current signals. Considering these different designs, the three sensing points in our architecture in Fig.~\ref{sensing_points} offer numerous potential sensing signals, ten of which we record and analyze as shown in Fig.~\ref{sensing_harware}.
There is no current flowing in the open circuit design, so we only record the voltage before and after the rectifier.
The voltage at sensing point 2b is the same as at 3b, therefore we do not sample it twice in Fig.~\ref{sensing_harware}b.
Lastly, we can not sample the AC current in converter-less and converter-based designs and the AC transducer voltage in the converter-based design due to hardware limitations.

We use a S230-J1FR-1808XB two-layered piezoelectric bending transducer from MID\'E technology\footnote{https://www.mide.com/} in all hardware designs.
All signals are sampled with a 12-bit \ac{adc} at a sampling frequency of \SI{100}{\hertz}.
The first design (see Fig.~\ref{sensing_harware}a) represents the open-circuit configuration and serves as a benchmark for comparison to the state of the art~\cite{khalifa2015energy, khalifa2017harke, umetsu2019ehaas, 8730510}.
The other two designs use a \SI{220}{\micro\farad} capacitor to temporarily store the harvested energy and an intermittently powered load\footnote{In this experiment, we set $V_{thr}^{on} = 3.38 V$ and $V_{thr}^{off} = 2.18 V$} consisting of two \acp{led}, mimicking the behaviour of a batteryless device. The third design (in Fig.~\ref{sensing_harware}c) uses a TI BQ25504 DC-DC boost-converter between the rectifier and the capacitor.
The converter is configured to regulate its input voltage to around \SI{1}{\volt}.
The prototypes are designed as dataloggers with a focus on enabling accurate measurements of signals at all sensing points rather than on optimizing harvesting efficiency.

In order to analyze the characteristics of the available signals, in an initial study, we collect data from a full-sized adult tricycle using the four hardware prototypes as shown in Fig.~\ref{fig:experiment_setup_bike}.
For data collection, the prototypes are placed in a plastic box with a size of \SI{39}{\centi\metre}$\times$\SI{29}{\centi\metre}$\times$\SI{17}{\centi\metre} and the piezoelectric transducers are mounted on a \SI{23}{\centi\metre} long metallic bar mounted inside the box.
We use a block tip mass of \SI{24.62}{\gram}$\pm$0.5\% with each transducer to make it more sensitive to lower frequency vibrations.
During the experiment using the tricycle, the box containing the prototypes is placed in the wire basket behind the saddle as shown in Fig.~\ref{fig:experiment_setup_bike}.
\begin{figure}[t!]
\centering
\includegraphics[width=9.5cm, height=16cm]{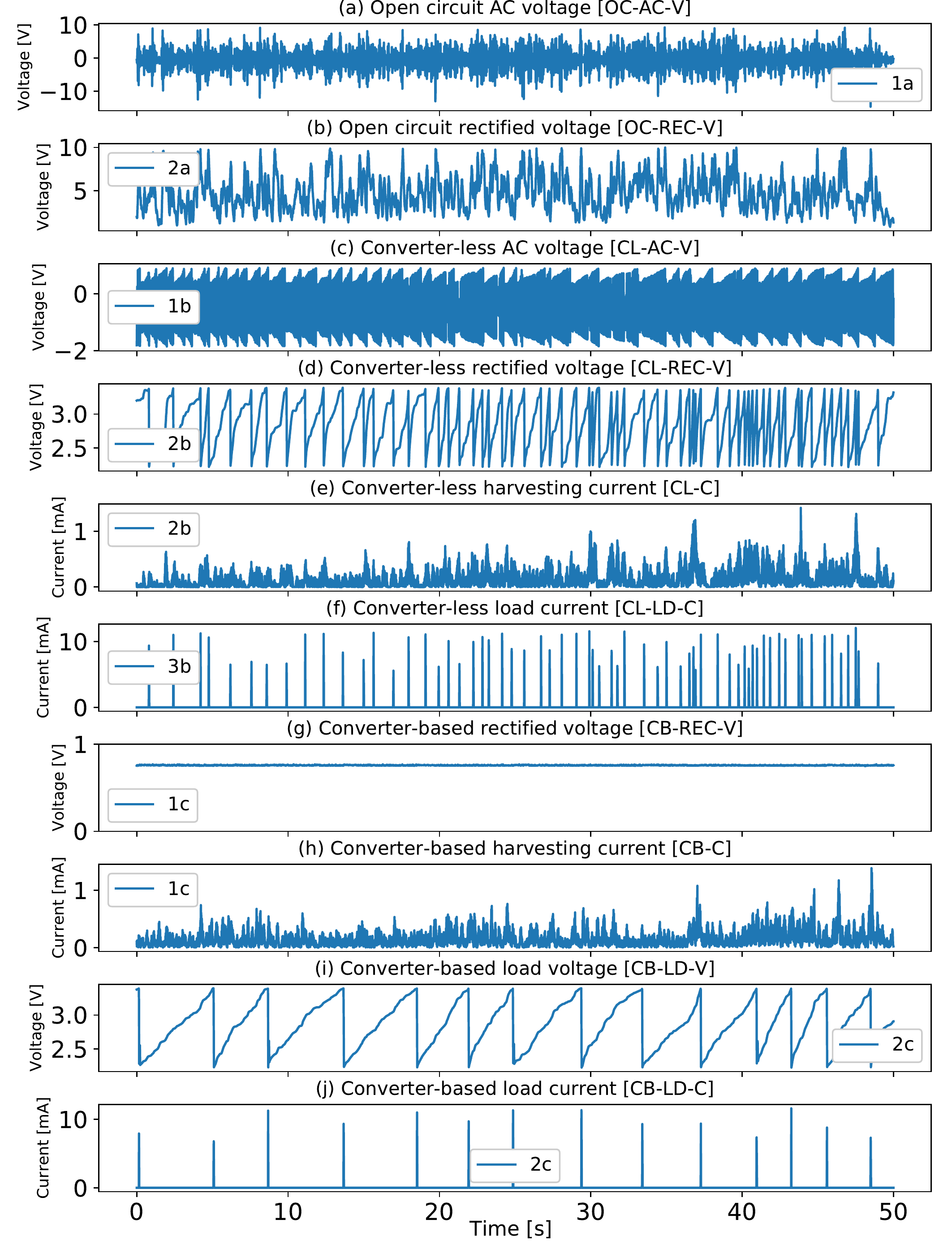}
\caption{Different types of signals from \ac{keh} on a tricycle; the labels indicate the sensing points in Fig.~\ref{sensing_harware}}
\label{fig:keh_signals_bike}
\vspace{-0.5cm}
\end{figure}

\subsection{The interference problem at different sensing points}
\label{impact_of_energy_harvesting}

Most previous works on \ac{keh}-based sensing use the open circuit AC voltage of the transducer as a sensing signal~\cite{khalifa2015energy, khalifa2017harke, 8730510, umetsu2019ehaas}. However, Fig.~\ref{fig:keh_signals_bike} depicts example data traces from all signals collected from our hardware prototypes. The AC voltage shown in Fig.~\ref{fig:keh_signals_bike}a is proportional to the displacement of the tip and thus accurately reflects the excitation of the transducer.
Fig.~\ref{fig:keh_signals_bike}b depicts the voltage after rectification, corresponding to the absolute values of the AC voltage.
When closing the circuit by connecting a capacitor to the output of the rectifier, the voltage on the transducer is enveloped by the voltage on the capacitor as discussed in Sec.~\ref{sec:batteryless_design}.
This has previously been described as the interference problem~\cite{ma2018sehs}.
In this section, we extend the discussion of the interference problem to the effect of a transiently powered load, consuming energy from the capacitor.
Furthermore, we explore various other sensing signals in different energy harvesting designs and discuss how they are affected by the capacitor and load.
\begin{figure}[t!]
\centering
\includegraphics[width=9cm, height=2.75cm]{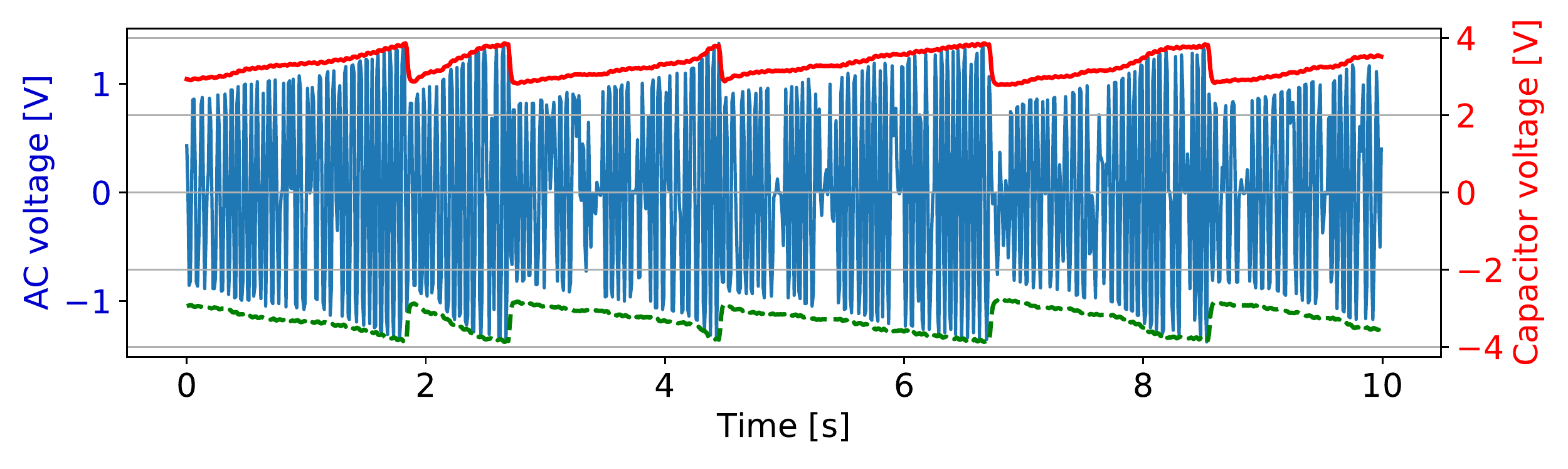}\vspace{-0.3cm}
\caption{\ac{keh} AC voltage is enveloped in the capacitor voltage}
\label{fig:envelope}
\vspace{-0.65cm}
\end{figure}

\fakepar{Effect of capacitor and load}
The effect of the combination of capacitor and load is illustrated in Fig.~\ref{fig:envelope}.
As expected, the AC voltage is enveloped by the capacitor voltage minus two diode drops (see Eq.~\ref{eq:envel_formation}).
The dashed green graph shows the mirrored capacitor voltage, highlighting that, due to the rectifier, the signal is enveloped in either polarity.
In a converter-less design, the capacitor is connected directly to the output of the rectifier.
Thus, the converter-less rectified voltage shown in Fig.~\ref{fig:keh_signals_bike}d equals the capacitor voltage.
The shape of the envelope is crucially determined by the behaviour of the transiently powered load that is illustrated in Fig.~\ref{fig:transiently_powered_sensors}.
The gradual rise reflects the charging process and the sharp drop in voltage is due to the load being switched on.
Nevertheless, the affected signals may still contain enough information to distinguish between various contexts with reasonable accuracy as described in Section~\ref{results} in detail.
The previous approach of applying filters to compensate for the capacitor charging curve~\cite{ma2018sehs} can not easily be applied when considering the hard-to-predict load behaviour.
Instead, we seek to explore different sensing signals in the energy harvesting circuit that may be less affected by the interference problem.

\begin{figure}[t!]
\centering
\includegraphics[width=9cm, height=5.5cm]{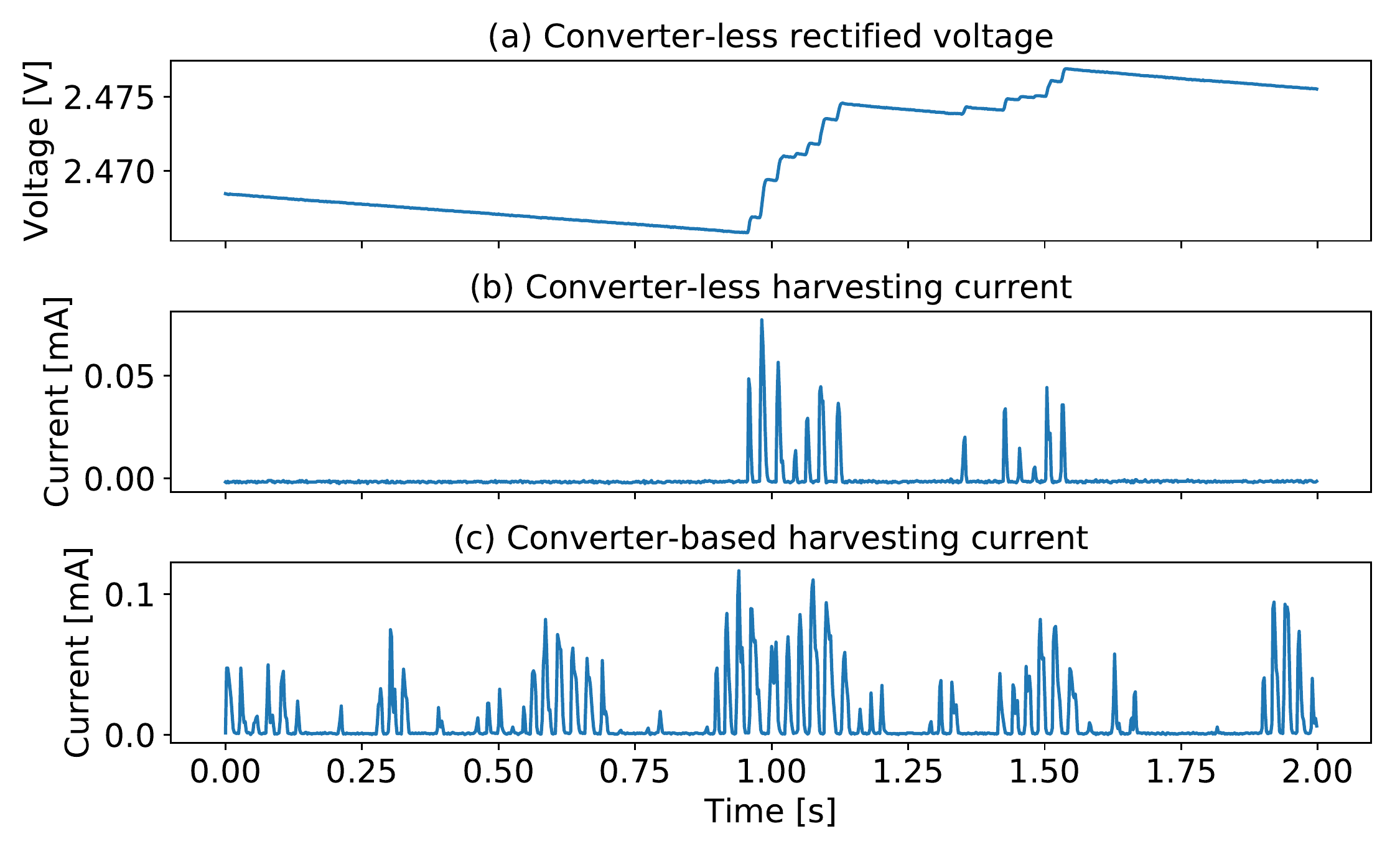}\vspace{-0.3cm}
\caption{Effect of capacitor voltage on the harvesting current}
\label{fig_interference_effect}
\vspace{-0.65cm}
\end{figure}

\fakepar{Current signals}
Fig.~\ref{fig_interference_effect} shows the harvesting current for converter-less and converter-based designs compared to the rectified harvesting voltage.
The rectified voltage equals the capacitor voltage and thus exhibits the expected \textit{envelope distortion}.
The current signals, in contrast, exhibit the typical waveform of a damped spring mass oscillator, which is a widely used model for a tip-mass loaded piezoelectric transducer~\cite{khalifa2017harke}.
They are approximately proportional to the displacement of the tip mass and thus incorporate details of the underlying physical process.
However, current can only flow, once the voltage across the transducer exceeds the voltage on the output of the rectifier plus two diode drops.
Thus, the current is zero when the voltage on the transducer is lower than the \textit{threshold voltage} and the corresponding information is lost.
We call this the \textit{threshold distortion}.
In a converter-less system, the threshold voltage is the varying capacitor voltage.
In a converter-based system, it is the constant, configurable input voltage of the DC-DC converter.
For example, in our measurement campaign, we empirically set the input voltage to \SI{1}{\volt}.
As a result, current starts to flow at lower voltages, resulting in higher energy yield and more context information in the converter-based current signal (Fig.~\ref{fig_interference_effect}c) than in the converter-less signal (Fig.~\ref{fig_interference_effect}b).
Threshold distortion is less critical than envelope distortion as it only affects the part of the signal when the amplitude of the signal is too low for harvesting.
For \textit{energy positive}, batteryless sensing, the transducer must anyway be dimensioned to provide sufficient energy when we want to sample the signal.
The envelope distortion instead affects the parts of the voltage signal when energy can be harvested.
In summary, the harvesting current signal is an attractive signal whose sensing potential has not been previously explored in sufficient detail.
\begin{table}[ht]
    \centering
    \caption{Detail of the collected data using \ac{keh}}
    \begin{tabular}{cccc}
    \toprule
       Transport & Duration with & Duration without & Number of\\
       mode & stops (minutes) & stops (minutes) & trials\\
       \midrule
       \midrule
       Ferry & 156 & 93 & 4\\
       Train & 125 & 96 & 4\\
       Bus & 144 & 76 & 6\\
       Car & 98 & 83 & 2 \\
       Tricycle & 72 & 59 & 4\\
       Pedestrian & 216 & 66 & 8\\
       \textbf{Total} & \textbf{811} & \textbf{473} & --\\
       \bottomrule
    \end{tabular}
    \label{tab:data_collection}
    \vspace{-0.5cm}
\end{table}
\section{Transport Mode Detection}
\label{Transport_Mode_Detection_Algorithm}
We employ \ac{tmd} as a case study to compare the sensing performance of the considered \ac{keh} signals. 

\subsection{Data collection}
Volunteers\footnote{Ethical approval has been granted from CSIRO for carrying out this experiment (Approval number 106/19)} are asked to carry the box (containing the hardware prototypes) while travelling using multiple transport modes including ferry, train, bus, car, tricycle and pedestrian movement.
During transitions between the vehicular transport modes, the volunteers walk as pedestrians, including slow walking, brisk walking, moving upstairs/downstairs and some stop periods. 
The data is collected from various transport modes in Brisbane city with variations in seating location, time of the day, subjects and transport routes, with an average duration of 78 minutes from each transport mode as shown in Table~\ref{tab:data_collection}. 
For each trial, we use a different vehicle or choose another route.
We also alternate the location of the prototypes within the corresponding vehicle (i.e., front, middle or rear section).
This ensures that the collected data is representative of all types of vibrations experienced in the vehicles.
In order to compare the performance of the proposed \ac{keh}-based architecture with the state-of-the-art, we also collect data using an MPU9250 3-axis accelerometer.
Finally, the volunteers manually record the actual transport mode across the course of the experiments, to serve as ground truth for classification.
We do not use load voltage and current (at points 3b and 2c in Fig.~\ref{sensing_harware}) for sensing as the load is not turned on frequently especially when vibrations are low such as on the ferry or the train.
Similarly, the rectified voltage in Fig.~\ref{sensing_harware}d (at point 1c) is also not used for extracting information as it is fixed to a specific level by the converter and does not contain information.
The remaining five \ac{keh} signals are analyzed in terms of their sensing potential in Section~\ref{results}.
\begin{figure}[t!]
\centering
\includegraphics[width=8cm, height=3.5cm]{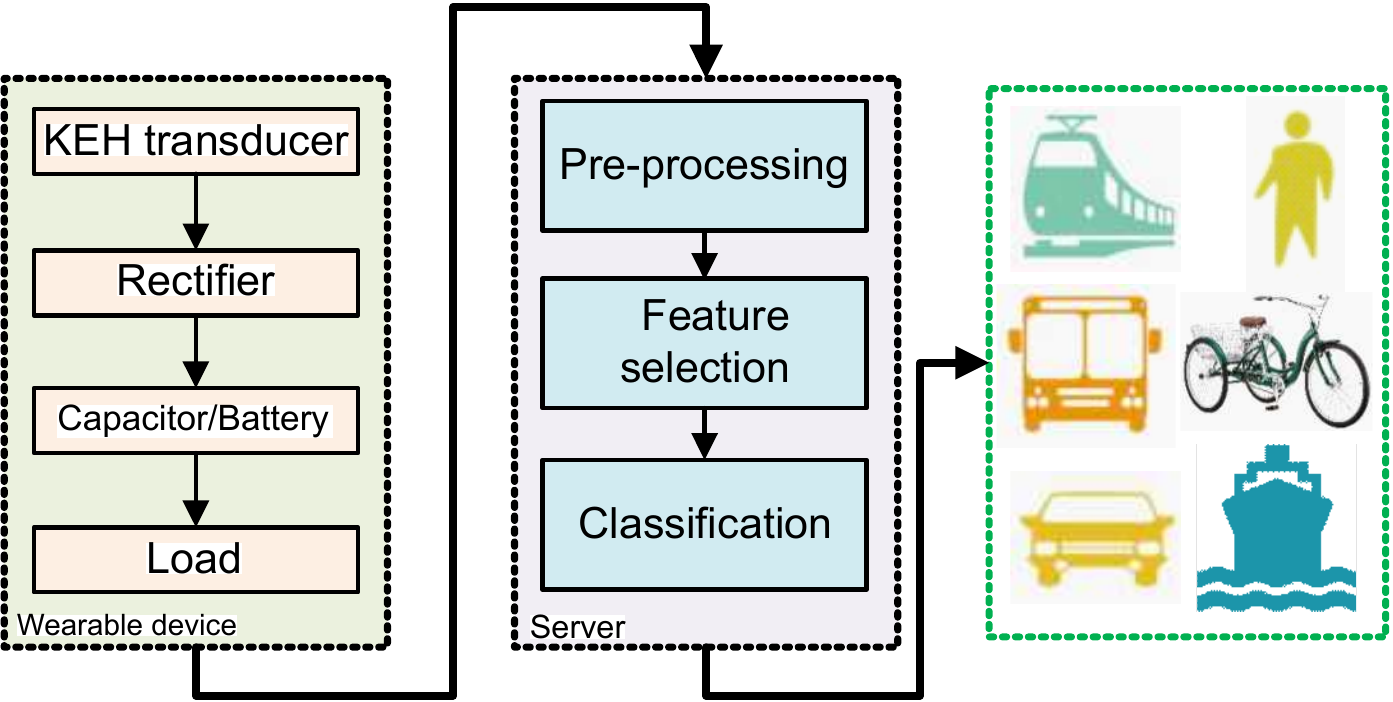}
\caption{The model architecture for transport mode detection}
\label{fig_sensing_model}
\vspace{-0.65cm}
\end{figure}
\begin{table}[ht]
    \centering
    \caption{The number of features selected using RFE}
    \begin{tabular}{cccc}
    \toprule
        Signal & Initial features & Using RFE & \acs{cv} score\\
        \midrule
        \midrule
         Accelerometer (Acc) & 128 & 36 & 0.95\\
         OC-AC-V & 42 & 26 & 0.93\\
         OC-REC-V & 42 & 31 & 0.86\\
         CL-AC-V & 42 & 26 & 0.92\\
         CL-REC-V & 42 & 36 & 0.84\\
         CL-C & 42 & 29 & 0.72\\
         CB-C & 42 & 09 & 0.93\\
         \bottomrule
    \end{tabular}
    \label{tab:features_overall}
    \vspace{-0.25cm}
\end{table}
\subsection{System model}
The proposed system model for \ac{tmd} is depicted in Fig.~\ref{fig_sensing_model}.
A wearable device collects real-time data from various sensing points in the \ac{keh} circuit while travelling on various transport modes.
A server processes the collected data, removes stops/pauses, extracts the dominant feature set for \ac{tmd}, and implements the classification techniques to classify the current transport mode. A computational unit located within the vehicle, along the travel route or a personal digital assistant can serve the purpose of a server. 
Below, we explain each component in detail. 

\begin{figure}[t]
\centering
\includegraphics[width=9cm, height=4cm]{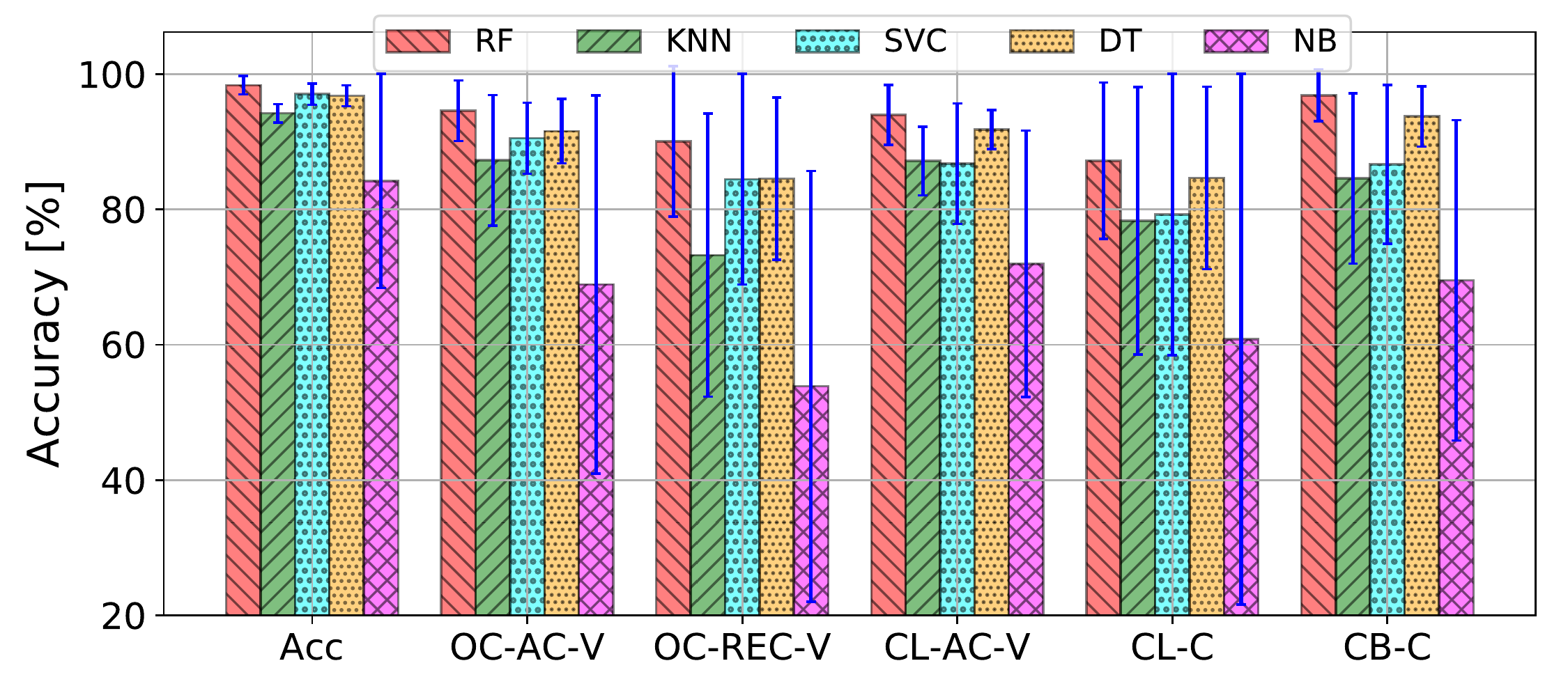}\vspace{-0.3cm}
\caption{\ac{tmd} accuracy of accelerometer and \ac{keh} signals using five classification algorithms (window size = 1 second)}
\label{fig:all_classifiers}
\vspace{-0.40cm}
\end{figure}
\subsubsection{Pre-processing}
First, we convert the \ac{adc} readings into actual voltage and current signals.
When the vehicle is stationary, for example at traffic lights, there are lower vibrations as compared to the moving state, making \ac{tmd} difficult~\cite{8730510}.
Therefore, we detect and remove these stops/pauses based on the average value of the signal~\cite{stockx2014subwayps}.


\subsubsection{Feature selection}
We divide the collected data into equal sized windows with a 50\% overlap and extract multiple time and frequency domain features as described in~\cite{hemminki2013accelerometer, khalifa2017harke, 8730510}.
As individual features can embed varying levels of information content about the transport mode, we employ \ac{rfe} to find the minimal and most significant feature set using \ac{cv}~\cite{guyon2002gene}.
Table~\ref{tab:features_overall} shows the total number of features selected in various types of sensing signals with the corresponding CV score.
It also depicts that only 9 features are selected out of 42 time and frequency domain features from the converter-based current signal.
These are the common features selected from all types of signals (which include maximum value, minimum value, amplitude range, coefficient of variation, skewness, kurtosis, inter-quartile range, absolute area, and root mean square value).
Note that the converter-based current signal offers comparable \ac{cv} score as the accelerometer and open circuit AC voltage signals using a smaller feature set, indicating the rich information content embedded in it. 

\subsubsection{Classification}
Five well-known machine learning classifiers are implemented including \ac{rf}, \ac{dt}, \ac{svm}, \ac{knn} and \ac{nb}. For each classifier, we perform 10-fold cross validation and plot all results averaged with 95\% confidence interval. Prior to the implementation of classification algorithms, we use \ac{smote}~\cite{chawla2002smote} to handle imbalanced data from the various transport modes and normalise the selected features with zero mean and standard deviation of one.
\section{Performance Evaluation}
\label{results}
In this section, we evaluate the performance of the proposed architecture using \ac{tmd} as a case study. 
\begin{figure}[ht]
\centering
\includegraphics[width=8.5cm, height=5cm]{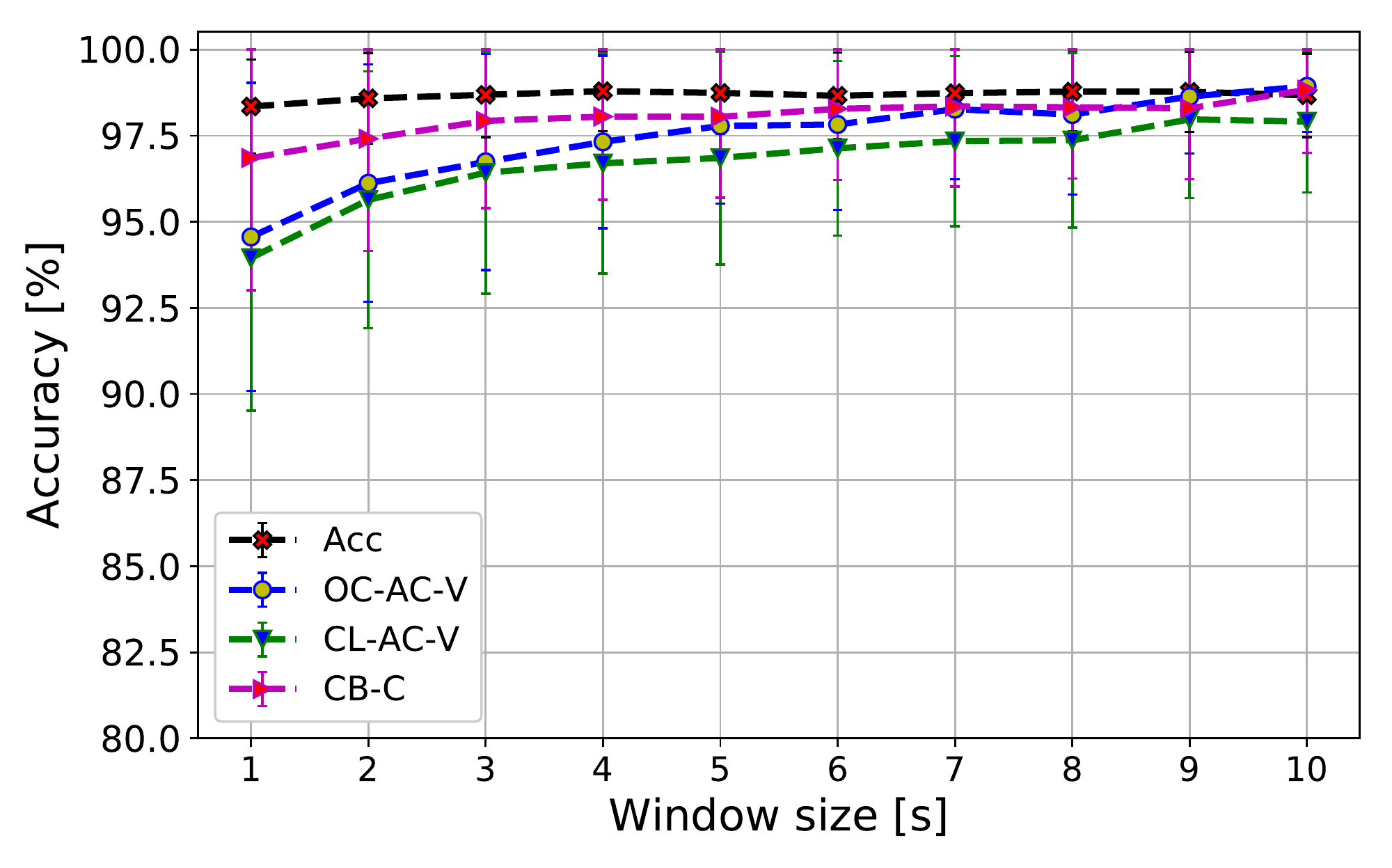}\vspace{-0.3cm}
\caption{Accuracy of \ac{tmd} using accelerometer and \ac{keh} signals with growing window sizes}
\label{fig:Transport_mode_detection_accuracy}
\vspace{-0.40cm}
\end{figure}
\subsection{Detection accuracy of \ac{keh}-based sensing signals}
\label{results_a}
\subsubsection{Different classifiers}
We compare \ac{tmd} accuracies of accelerometer and five \ac{keh}-based sensing signals using the considered classification schemes in Fig.~\ref{fig:all_classifiers}.
The results indicate that the \ac{rf} classifier outperforms other classifiers for all signals including accelerometer.
Therefore, in the rest of the document, we present the results from the \ac{rf} classifier.
For all classifiers, the open circuit rectified voltage achieves lower detection accuracy than the open circuit AC voltage due to the loss of information by inverting the negative part of the signal.
Also, the converter-less current signal achieves lower detection accuracy than the converter-based current signal due to the interference problem of the capacitor voltage in the former as discussed in Section~\ref{impact_of_energy_harvesting}.
Intuitively, although the pattern of converter-less AC voltage is affected due to the interference problem shown in Fig.~\ref{fig:envelope}, we find that its context detection performance (93.95\%) is not highly affected compared to the open circuit AC voltage (94.56\%); however, it provides lower detection accuracy than the accelerometer signal (98.35\%) as depicted in Fig.~\ref{fig:all_classifiers}. Converter-based current signal offers the highest detection accuracy (96.85\%) among all \ac{keh}-based signals which is close to the detection accuracy of the accelerometer signal (98.35\%) for a window size of one second.
This shows that using the converter to decouple the transducer and capacitor, allows the flow of current during smaller vibrations, ultimately yielding context-rich \ac{keh} signals. It is worth mentioning that any single axis signal of accelerometer offers lower detection accuracy ($Acc_x$: 92.10\%, $Acc_y$: 93.12\%, $Acc_z$: 90.70\%) than the combined 3-axis signal (98.35\%), for a window size of one second, due to the higher information content across spatial dimensions embedded into the 3-axis signal. Therefore, all results presented in this document employ a 3-axis accelerometer signal as the key benchmark, while \ac{keh} signals are single axis in nature.

\subsubsection{Varying window size} 
We now study the impact of window size on the \ac{tmd} accuracy of accelerometer, open circuit AC voltage, converter-less AC voltage and converter-based current, as the four signals that provide highest detection accuracy.
We plot the accuracy for each signal with varying window sizes from \SI{1}{\second} to \SI{10}{\second} in Fig.~\ref{fig:Transport_mode_detection_accuracy}.
It is clearly shown that the \ac{tmd} accuracy increases with the increasing window sizes for all including the accelerometer and KEH signals.
Although the open circuit AC voltage signal offers lower accuracy than the accelerometer signal for smaller window sizes, it provides comparable accuracy to the accelerometer signal for window sizes greater than \SI{7}{\second}.
Similarly, with converter-less AC voltage, the detection accuracy increases with the increasing window size as depicted in Fig.~\ref{fig:Transport_mode_detection_accuracy}.
However, it is slightly lower than the open circuit AC voltage for all window sizes due to the interference problem experienced with the inclusion of energy harvesting circuit.
Nevertheless, the converter-based current signal achieves comparable accuracy to the accelerometer even with a smaller window size (starting from \SI{1}{\second}) as shown in Fig.~\ref{fig:Transport_mode_detection_accuracy}.
Furthermore, for larger window sizes (\SI{10}{\second}), the converter-based current signal provides the same accuracy as that of the accelerometer signal (98.5\%).
It is worth noting that the \ac{keh} converter-based current signal offers high detection accuracy using less features than the conventional 3-axis accelerometer signal as depicted in Table~\ref{tab:features_overall}.

\begin{figure}[t!]
\centering
\includegraphics[width=9cm, height=4.5cm]{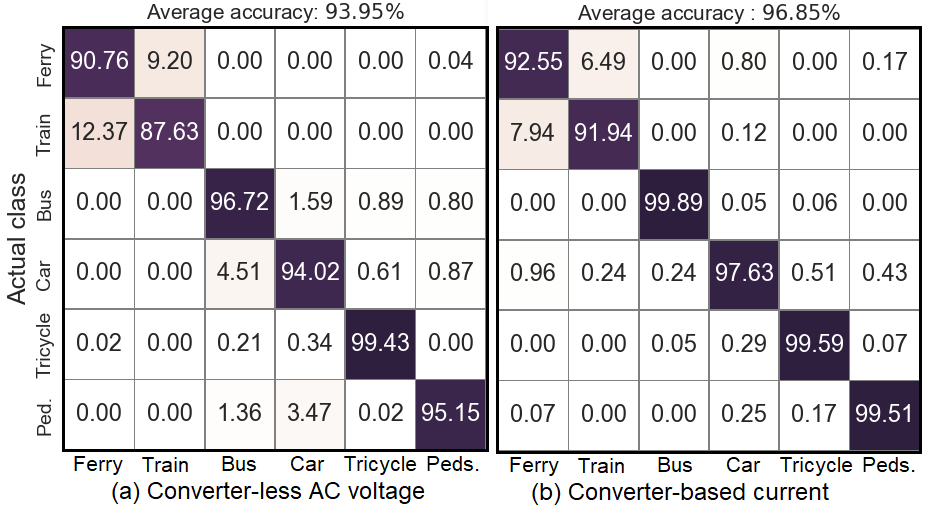}\vspace{-0.2cm}
\caption{Confusion matrices for \ac{tmd} using (a) \ac{keh} converter-less AC voltage and (b) \ac{keh} converter-based current signals (window size = 1 second)}
\label{Confusion_matrices}
\vspace{-0.65cm}
\end{figure}
\subsubsection{Current signal vs AC voltage}
We now focus on comparing the \ac{keh} current signal with the converter-less AC voltage signal in order to distinguish the considered transport modes.
Fig.~\ref{Confusion_matrices} shows the confusion matrices of both signals, highlighting that converter-less AC voltage signal provides detection accuracy of more than 94\% for most of the transport modes. However, the detection accuracy for trains (87.63\%) and ferries (90.76\%) is the lowest as both of these transport modes have dedicated non-road transport surfaces (i.e., river  water  and  tracks  respectively), resulting in similar, low vibration amplitudes.
Furthermore, the lower signal amplitude for these transport modes is more susceptible to noise which makes it harder to differentiate between them.
Similarly, for converter-based current signal, an accuracy of higher than 97\% is achieved for most of the transport modes except for ferries and trains where 92.55\% and 91.94\% accuracies are achieved, respectively.
The reason behind the high detection accuracy of the converter-based current lies in the lower threshold distortion as compared to the converter-less signals (voltage and current) as depicted in Fig.~\ref{fig_interference_effect}.
Other types of signals (like converter-less AC voltage and current) are affected by the charged capacitor as it hinders the flow of current from the transducer especially during lower vibrations. On the other hand, in the converter-based design, the capacitor voltage is decoupled from the transducer voltage and current flows even during lower vibrations which reflects the detailed physical phenomenon and enhances the detection accuracy.
\begin{table}[t!]
    \centering
    \caption{Harvested power for various transport modes using converter-less and converter-based \ac{keh} circuits}
    \begin{tabular}{ccc}
    \toprule
        \multirow{2}{*}{Transport mode} & \multicolumn{2}{c}{Harvested Power [\si{\micro\watt}]}\\
         & Converter-less & Converter-based\\ \midrule
        \midrule
         Ferry & 0.1 & 0.2  \\
         \midrule
         Train & 0.06 & 0.1  \\
         \midrule
         Bus & 3.2 & 20.4  \\
         \midrule
         Car & 4.0 & 27.8  \\
         \midrule
         Tricycle & 5.3  & 20.4 \\
         \midrule
         Pedestrian & 3.7  & 10.5  \\
         \midrule
         \textbf{Average} & \textbf{2.7} & \textbf{13.2} \\
         
         \bottomrule
    \end{tabular}
    \label{tab:harvested_energy}
    \vspace{-0.35cm}
\end{table}
\subsection{Energy harvesting}
\label{Energy_Harvesting_using_keh}

In order to compare the energy harvesting potential between transport modes and circuit designs, we calculate the average harvesting power by first calculating the stored energy on the capacitor for any point in time, then adding up all positive changes in stored energy and finally dividing by the recording duration.
The results are presented in Table~\ref{tab:harvested_energy}.
The highest power is harvested in the car and on the tricycle, where we observe strong vibrations close to the resonant frequency of the transducer ($\approx$\SI{25}{\hertz}).
On the other hand, due to smooth pathways and lower vibration amplitudes, we record the lowest power on the ferry and the train.
In most cases, we observe significantly higher energy yield with the converter-based design than with the converter-less energy harvesting design.
\subsection{Energy consumption and system costs}
\label{Power_Consumption_Analysis}

\begin{figure}[t!]
\centering
\includegraphics[width=6cm, height=3cm]{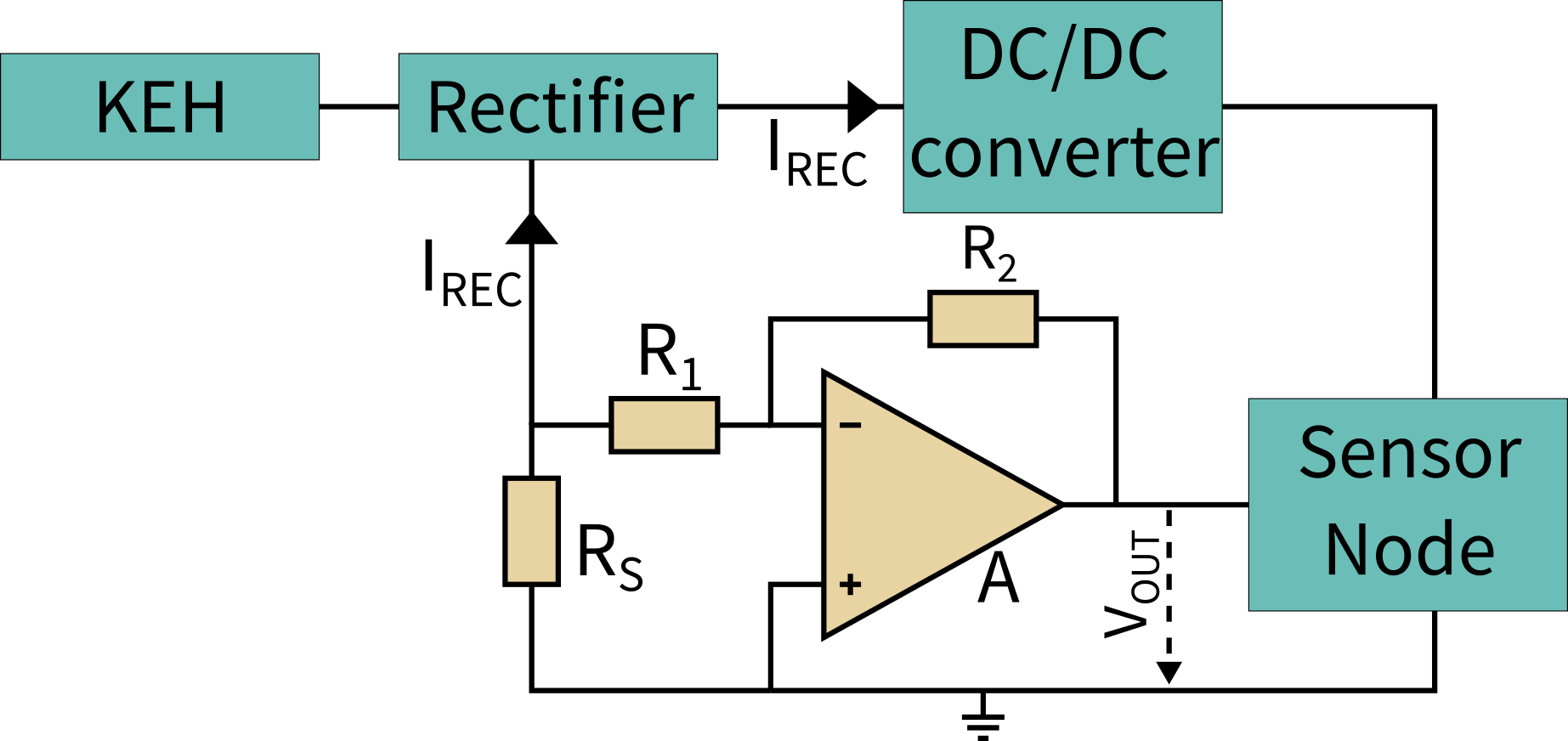}
\caption{Current sensing mechanism using an amplifier}
\label{fig:keh_current_sensing_amplifier}
\vspace{-0.65cm}
\end{figure}

In Section~\ref{results_a}, we discovered that the current into the DC-DC converter can be used as a sensing signal for \ac{tmd}, achieving detection accuracy comparable to a 3-axis accelerometer signal.
Previous work found that sampling the voltage of a \ac{keh} transducer consumes orders of magnitude less energy than sampling a state-of-the-art accelerometer~\cite{khalifa2017harke}.
In this section, we show how the harvesting current signal can be accessed in a real system and how this solution compares to the accelerometers in terms of power requirements and costs.
We propose to use a shunt ampere-meter to convert the current into a voltage that can easily be sampled by the sensor node using an \ac{adc} as shown in Fig.~\ref{fig:keh_current_sensing_amplifier}.
The current over resistor $R_S$ causes a negative voltage drop that is inverted and amplified by the inverting operational amplifier $A$.
By choosing the corresponding resistor values, the amplifier output can be adjusted to match the input range of the \ac{adc}.

For all power calculations, we assume a supply voltage of \SI{3}{\volt}.
Using a low power operational amplifier (e.g. TI LPV521, \SI{350}{\nano\ampere}) and low value for the shunt resistor, the sum of power consumption and losses are around \SI{2}{\micro\watt} under typical harvesting conditions.
This is more than two orders of magnitude less than the power consumption of the lowest power analog accelerometer (ADXL356, \SI{450}{\micro\watt}) that we could find.

When adding the current for an external, low power \ac{adc} (e.g., ADS7042, $\approx$\SI{700}{\nano\watt}@\SI{100}{\hertz}), the power consumption of the proposed system is still less than \SI{3}{\micro\watt}.
Highly integrated, digital accelerometers achieve significantly lower power consumption than their analog counterparts (e.g., ADXL363, \SI{7.2}{\micro\watt}@\SI{100}{\hertz}) by co-design of the sensor and the signal acquisition chain and duty-cycling according to the configured sampling rate.
This is still more than twice as much as our proposed \ac{keh} current sensor consumes based on discrete, off the shelf components.
We expect that integrating our circuit with an optimized signal acquisition chain would further reduce the power consumption.

\begin{figure}[ht]
\vspace{-0.5cm}
\centering
\includegraphics[width=8cm, height=3cm]{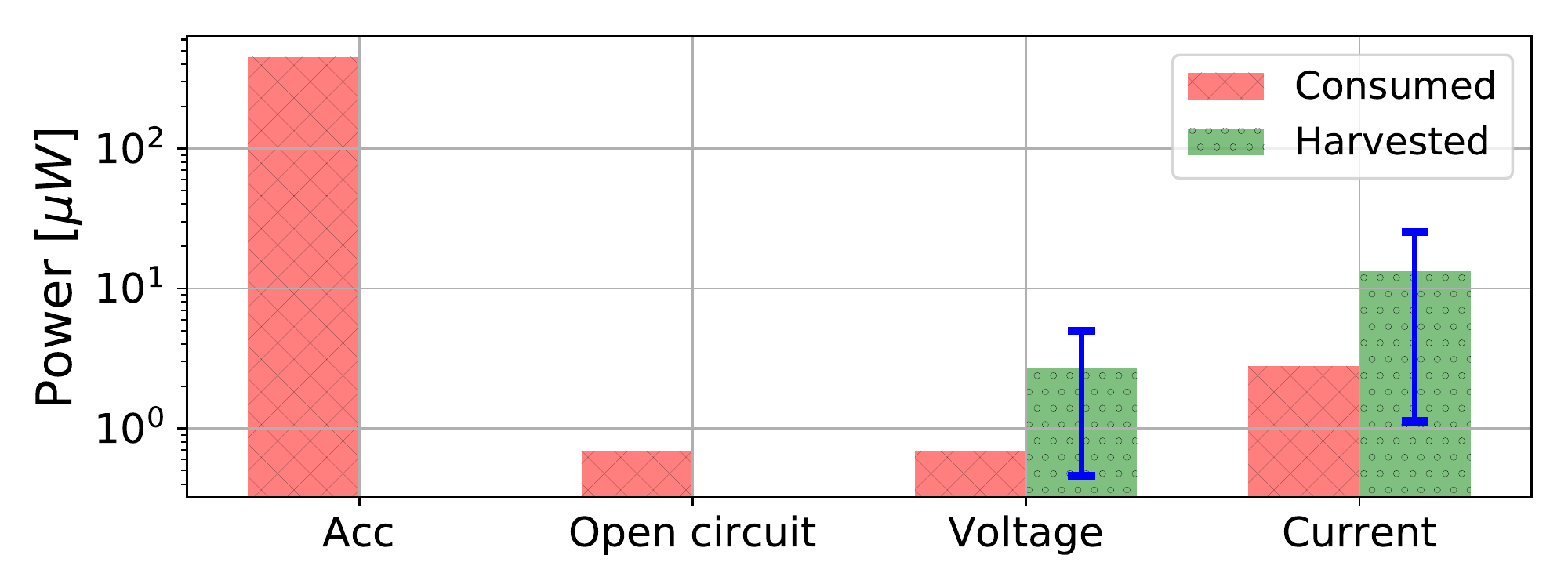}\vspace{-0.3cm}
\caption{Comparison between the consumed power in signal acquisition and the harvested power using \ac{keh}}
\label{fig:power_consumption_analysis}
\vspace{-0.45cm}
\end{figure}

Fig.~\ref{fig:power_consumption_analysis} compares the harvested and consumed power for accelerometer, open circuit voltage, converter-less voltage and converter-based current.
It shows that the \ac{keh} circuits consume significantly lower energy than the accelerometer for providing the sensing signals.
The accelerometer and \ac{keh} open circuit design consume power from an external source without generating energy.
On the other hand, on average, converter-less and converter-based designs harvest more energy than the required for signal acquisition leading towards \textit{energy positive sensing}.


\fakepar{Costs}
The component costs for the proposed \ac{keh} current sensing circuit, including the amplifier (0.49 USD) and three resistors ($<$0.01 USD) per device in quantities of 1000 is around 0.50 USD. This is less than a third of the price of the cheapest accelerometer in the same quantity (Kionix Inc. KXTC9-2050-FR, 1.54 USD) that we could find.

\subsection{Energy positive sensing: Discussion and analysis}

\begin{figure}[t!]
\centering
\includegraphics[width=8.5cm, height=6cm]{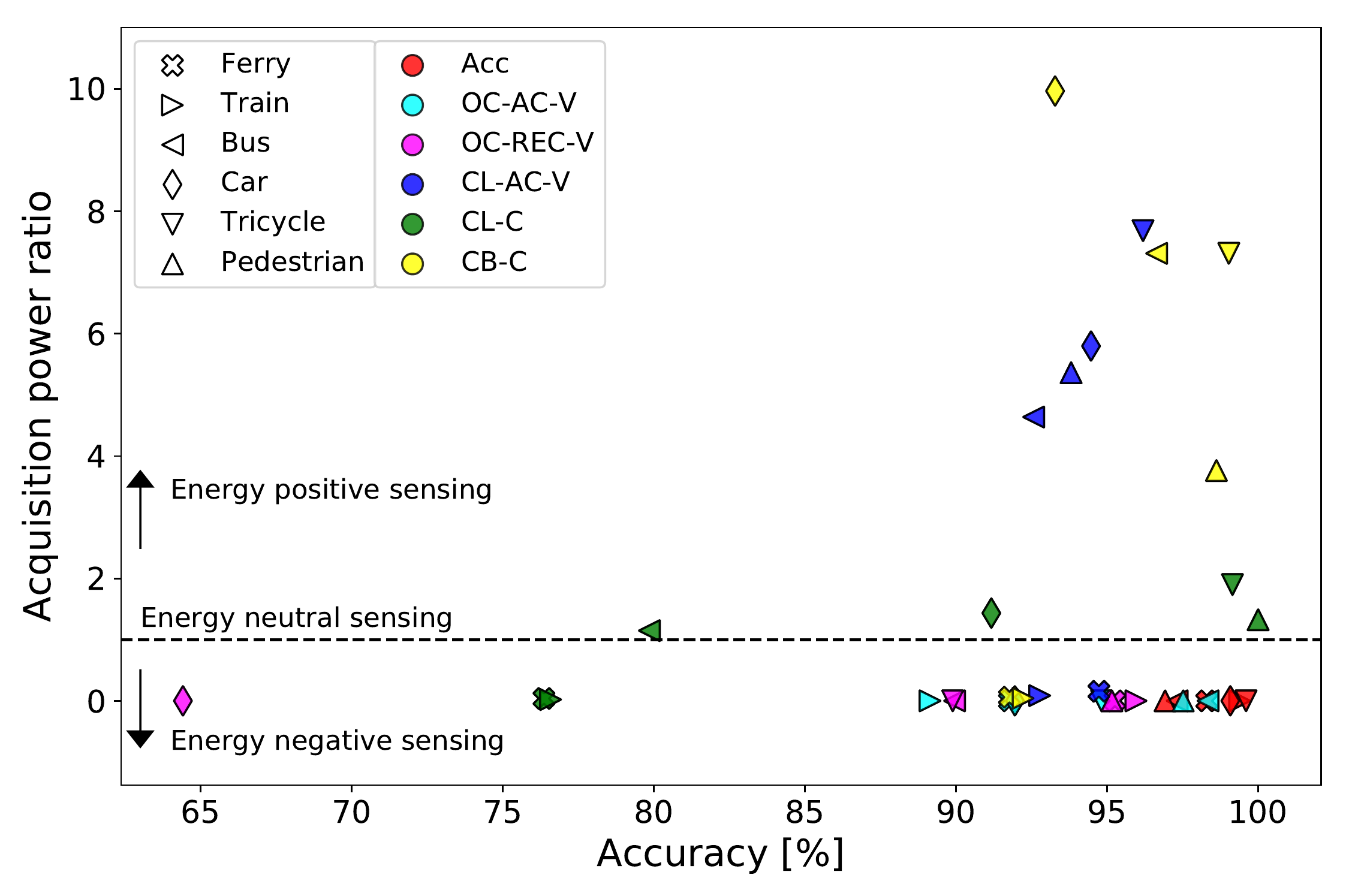}\vspace{-0.3cm}
\caption{Comparison of various signals in terms of \ac{apr} and \ac{tmd} accuracy (window size = 1 second)}
\label{overall_bar_plot}
\vspace{-0.65cm}
\end{figure}

In order to analyze \textit{energy positive sensing} quantitatively, we define the \ac{apr} as the ratio of harvested power ($P_{har}$) and power required for signal acquisition ($P_{acq}$). 
\begin{equation}
    APR = \frac{P_{har}}{P_{acq}}
\end{equation}
Energy negative sensing has \ac{apr}$\,<\,$1 and \textit{energy positive} has \ac{apr}$\,>\,$1.
The boundary between the two classes at \ac{apr}\,=\,1 represents \textit{energy neutral sensing}. In Fig.~\ref{overall_bar_plot}, we plot the \ac{apr} over the achieved accuracy of \ac{tmd} for all combinations of transport modes and sensing signals.
Although the accelerometer signal provides the highest detection accuracy, it has zero \ac{apr} due to zero harvested power.
Similarly, \ac{keh}-based sensing in an open circuit configuration offers lower accuracy than the accelerometer signal and provides zero \ac{apr} as no energy is being harvested. Instead both accelerometer and KEH open circuit designs consume energy from an external source for signal acquisition.
Therefore, both of these devices are energy negative for all transport modes.
The transducer voltage in the converter-less design (CL-AC-V) offers \textit{energy positive sensing} with \ac{apr} of four to eight for four out of six considered transport modes with reasonable detection accuracy.
However, the converter-based current signal (CB-C) outperforms all KEH signals and offers \ac{tmd} accuracy comparable to the accelerometer signal with an \ac{apr} of four to ten for four out of six considered transport modes.

In summary, the results show that \textit{energy positive sensing} is possible with both converter-less and converter-based energy harvesting designs for at least two-thirds of the considered transport modes.
Depending on the transport mode, \ac{keh}-based sensing circuit provides up to ten times as much power as required for the signal acquisition while offering detection accuracy close to the 3-axis accelerometer.

\section{Related Work}
\label{Literature_Review}
We categorize the literature into three classes based on the utilization of KEH for energy generation, sensing and simultaneous sensing and energy harvesting.
\subsection{KEH as a source of energy}

Many previous works employed \ac{keh} as a source of energy~\cite{magno2018micro,fan2018capturing} to replace the non-rechargeable batteries in \ac{iot} for their widespread deployment. The authors in~\cite{magno2018micro} presented a KEH wearable device which harvests energy from daily human activities. Whereas, Fan et al.~\cite{fan2018capturing} employed KEH in a shoe to harvest energy from the foot strikes during walking and running. However, still the harvested energy from a single tiny KEH is not enough to power the conventional sensors (such as accelerometers) continuously which pushes researchers to find alternate methods for perpetual and sustainable sensing.

\subsection{KEH as a sensor}
Recently, \ac{keh} is also used for human and machine context detection in various applications, reducing the conventional sensor related power consumption~\cite{khalifa2017harke}. The authors in~\cite{khalifa2017harke, kalantarian2015monitoring, xu2018keh, lin2019h2b} used \ac{keh} open circuit AC voltage for human activity recognition, monitoring eating habits, gait recognition and sensing heartbeat respectively. As the vibration pattern is distinct during different human activities like walking, running, standing, etc., the generated energy from \ac{keh} contains information about the type of activity being performed. Similarly, Lan et al.~\cite{lan2017capsense} employed a capacitor to store the harvested energy from \ac{keh} and then use the charging rate of the capacitor for human activity recognition. 
In order to extract rich context information, multi-source energy harvesters can also be employed. Umetsu et al.~\cite{umetsu2019ehaas} used \ac{seh} and \ac{keh} for room level location detection, where, SEH differentiates between the indoor and outdoor environments while \ac{keh} provides information about the type of human activity e.g., walking, sitting, standing, etc.

\subsection{Simultaneous sensing and energy harvesting KEH}


Building on the two aspects of KEH, i.e. sensing and energy generation, Ma et al.~\cite{ma2018sehs} proposed a technique to achieve both simultaneously for human gait recognition. However, they used two piezoelectric transducers in the front and rear of a shoe, where the second transducer significantly adds to the weight and costs of the system. Furthermore, their proposed system only considers the charging curve of the capacitor, neglecting the effects of a dynamic load consuming energy from the capacitor. It is thus not applicable in practical scenarios, where
the harvested energy is used to power the system. We address these issues by presenting a system architecture for \textit{energy positive sensing}, where a single transducer is used for sensing while simultaneously powering a dynamic load.

\section{Conclusion and Future Work}
\label{Conclusion_and_Future_Work}
In this paper, we present a scheme to use \ac{keh} simultaneously as a source for energy and information enabling energy positive sensing.
The novelty of this work lies in the exploration of voltage and current signals at various sensing points in the energy harvesting circuit.
In addition to sensing, we utilize the harvested energy from \ac{keh} to power a realistic load.
We present transport mode detection as a case study, design four different KEH prototypes and collect data from six transport systems.
Five classification techniques are implemented on five types of \ac{keh} signals and it is concluded that the harvesting current in a converter-based energy harvesting circuit achieves detection accuracy close to the accelerometer signal with two-fold lower energy consumption.
The results show that energy positive sensing is possible for at least two-thirds of the considered transport modes.
We expect that the proposed scheme triggers the exploration of energy positive sensing using other transducers as well. In future, multi-axis \ac{keh} sensing and energy harvesting can be explored to obtain even higher detection accuracy as well as harvested energy.


\bibliographystyle{IEEEtran}


\begin{thebibliography}{10}
\providecommand{\url}[1]{#1}
\csname url@samestyle\endcsname
\providecommand{\newblock}{\relax}
\providecommand{\bibinfo}[2]{#2}
\providecommand{\BIBentrySTDinterwordspacing}{\spaceskip=0pt\relax}
\providecommand{\BIBentryALTinterwordstretchfactor}{4}
\providecommand{\BIBentryALTinterwordspacing}{\spaceskip=\fontdimen2\font plus
\BIBentryALTinterwordstretchfactor\fontdimen3\font minus
  \fontdimen4\font\relax}
\providecommand{\BIBforeignlanguage}[2]{{%
\expandafter\ifx\csname l@#1\endcsname\relax
\typeout{** WARNING: IEEEtran.bst: No hyphenation pattern has been}%
\typeout{** loaded for the language `#1'. Using the pattern for}%
\typeout{** the default language instead.}%
\else
\language=\csname l@#1\endcsname
\fi
#2}}
\providecommand{\BIBdecl}{\relax}
\BIBdecl

\bibitem{cheol2018wearable}
I.~C~Jeong, D.~Bychkov, and P.~C. Searson, ``Wearable devices for precision
  medicine and health state monitoring,'' \emph{IEEE Transactions on Biomedical
  Engineering}, vol.~66, no.~5, pp. 1242--1258, 2018.

\bibitem{scalise2018wearables}
L.~Scalise and G.~Cosoli, ``Wearables for health and fitness: Measurement
  characteristics and accuracy,'' in \emph{2018 IEEE International
  Instrumentation and Measurement Technology Conference (I2MTC)}.\hskip 1em
  plus 0.5em minus 0.4em\relax IEEE, 2018, pp. 1--6.

\bibitem{hegde2017automatic}
N.~Hegde, M.~Bries, T.~Swibas, E.~Melanson, and E.~Sazonov, ``Automatic
  recognition of activities of daily living utilizing insole-based and
  wrist-worn wearable sensors,'' \emph{IEEE journal of biomedical and health
  informatics}, vol.~22, no.~4, pp. 979--988, 2017.

\bibitem{dias2018wearable}
D.~Dias and J.~Paulo Silva~Cunha, ``Wearable health devices--vital sign
  monitoring, systems and technologies,'' \emph{Sensors}, vol.~18, no.~8, p.
  2414, 2018.

\bibitem{castro2013taxi}
P.~S. Castro, D.~Zhang, C.~Chen, S.~Li, and G.~Pan, ``From taxi gps traces to
  social and community dynamics: A survey,'' \emph{ACM Computing Surveys
  (CSUR)}, vol.~46, no.~2, p.~17, 2013.

\bibitem{liu2014exploiting}
Y.~Liu, C.~Liu, N.~J. Yuan, L.~Duan, Y.~Fu, H.~Xiong, S.~Xu, and J.~Wu,
  ``Exploiting heterogeneous human mobility patterns for intelligent bus
  routing,'' in \emph{2014 IEEE International Conference on Data Mining}.\hskip
  1em plus 0.5em minus 0.4em\relax IEEE, 2014, pp. 360--369.

\bibitem{kumari2017increasing}
P.~Kumari, L.~Mathew, and P.~Syal, ``Increasing trend of wearables and
  multimodal interface for human activity monitoring: A review,''
  \emph{Biosensors and Bioelectronics}, vol.~90, pp. 298--307, 2017.

\bibitem{welch2018wearable}
K.~C. Welch, A.~S. Kulkarni, A.~M. Jimenez, and B.~Douglas, ``Wearable sensing
  devices for human-machine interaction systems,'' in \emph{2018 United States
  National Committee of URSI National Radio Science Meeting (USNC-URSI
  NRSM)}.\hskip 1em plus 0.5em minus 0.4em\relax IEEE, 2018, pp. 1--2.

\bibitem{taraldsen2012physical}
K.~Taraldsen, S.~F. Chastin, I.~I. Riphagen, B.~Vereijken, and J.~L. Helbostad,
  ``Physical activity monitoring by use of accelerometer-based body-worn
  sensors in older adults: a systematic literature review of current knowledge
  and applications,'' \emph{Maturitas}, vol.~71, no.~1, pp. 13--19, 2012.

\bibitem{hester2017future}
J.~Hester and J.~Sorber, ``The future of sensing is batteryless, intermittent,
  and awesome,'' in \emph{Proceedings of the 15th ACM Conference on Embedded
  Network Sensor Systems}.\hskip 1em plus 0.5em minus 0.4em\relax ACM, 2017,
  p.~21.

\bibitem{khalifa2017harke}
S.~Khalifa, G.~Lan, M.~Hassan, A.~Seneviratne, and S.~K. Das, ``Harke: Human
  activity recognition from kinetic energy harvesting data in wearable
  devices,'' \emph{IEEE Transactions on Mobile Computing}, vol.~17, no.~6, pp.
  1353--1368, 2017.

\bibitem{khalifa2015energy}
S.~Khalifa, M.~Hassan, A.~Seneviratne, and S.~K. Das, ``Energy-harvesting
  wearables for activity-aware services,'' \emph{IEEE internet computing},
  vol.~19, no.~5, pp. 8--16, 2015.

\bibitem{8730510}
G.~Lan, W.~Xu, D.~Ma, S.~Khalifa, M.~Hassan, and W.~Hu, ``Entrans: Leveraging
  kinetic energy harvesting signal for transportation mode detection,''
  \emph{IEEE Transactions on Intelligent Transportation Systems}, 2019.

\bibitem{umetsu2019ehaas}
Y.~Umetsu, Y.~Nakamura, Y.~Arakawa, M.~Fujimoto, and H.~Suwa, ``Ehaas: Energy
  harvesters as a sensor for place recognition on wearables,'' in
  \emph{Proceedings of the 2019 IEEE International Conference on Pervasive
  Computing Communications (PerCom)}.\hskip 1em plus 0.5em minus 0.4em\relax
  IEEE, 2019, pp. 1--10.

\bibitem{lan2017capsense}
G.~Lan, D.~Ma, W.~Xu, M.~Hassan, and W.~Hu, ``Capsense: Capacitor-based
  activity sensing for kinetic energy harvesting powered wearable devices,'' in
  \emph{Proceedings of the 14th EAI International Conference on Mobile and
  Ubiquitous Systems: Computing, Networking and Services}.\hskip 1em plus 0.5em
  minus 0.4em\relax ACM, 2017, pp. 106--115.

\bibitem{ma2018sehs}
D.~Ma, G.~Lan, W.~Xu, M.~Hassan, and W.~Hu, ``Sehs: Simultaneous energy
  harvesting and sensing using piezoelectric energy harvester,'' in \emph{2018
  IEEE/ACM Third International Conference on Internet-of-Things Design and
  Implementation (IoTDI)}.\hskip 1em plus 0.5em minus 0.4em\relax IEEE, 2018,
  pp. 201--212.

\bibitem{kalantarian2015monitoring}
H.~Kalantarian, N.~Alshurafa, T.~Le, and M.~Sarrafzadeh, ``Monitoring eating
  habits using a piezoelectric sensor-based necklace,'' \emph{Computers in
  biology and medicine}, vol.~58, pp. 46--55, 2015.

\bibitem{geissdoerfer2019preact}
K.~Geissdoerfer, B.~Kusy, R.~Jurdak, and M.~Zimmerling, ``Getting more out of
  energy-harvesting systems: Energy management under time-varying utility with
  preact,'' in \emph{2019 18th ACM/IEEE International Conference on Information
  Processing in Sensor Networks (IPSN)}.\hskip 1em plus 0.5em minus 0.4em\relax
  IEEE, 2019, pp. 109--120.

\bibitem{gomez2016dynamic}
A.~Gomez, L.~Sigrist, M.~Magno, L.~Benini, and L.~Thiele, ``Dynamic energy
  burst scaling for transiently powered systems,'' in \emph{Proceedings of the
  2016 Conference on Design, Automation \& Test in Europe}.\hskip 1em plus
  0.5em minus 0.4em\relax EDA Consortium, 2016, pp. 349--354.

\bibitem{stockx2014subwayps}
T.~Stockx, B.~Hecht, and J.~Sch{\"o}ning, ``Subwayps: towards smartphone
  positioning in underground public transportation systems,'' in
  \emph{Proceedings of the 22nd ACM SIGSPATIAL International Conference on
  Advances in Geographic Information Systems}.\hskip 1em plus 0.5em minus
  0.4em\relax ACM, 2014, pp. 93--102.

\bibitem{hemminki2013accelerometer}
S.~Hemminki, P.~Nurmi, and S.~Tarkoma, ``Accelerometer-based transportation
  mode detection on smartphones,'' in \emph{Proceedings of the 11th ACM
  conference on embedded networked sensor systems}.\hskip 1em plus 0.5em minus
  0.4em\relax ACM, 2013, p.~13.

\bibitem{guyon2002gene}
I.~Guyon, J.~Weston, S.~Barnhill, and V.~Vapnik, ``Gene selection for cancer
  classification using support vector machines,'' \emph{Machine learning},
  vol.~46, no. 1-3, pp. 389--422, 2002.

\bibitem{chawla2002smote}
N.~V. Chawla, K.~W. Bowyer, L.~O. Hall, and W.~P. Kegelmeyer, ``Smote:
  synthetic minority over-sampling technique,'' \emph{Journal of artificial
  intelligence research}, vol.~16, pp. 321--357, 2002.

\bibitem{magno2018micro}
M.~Magno, D.~Kneub{\"u}hler, P.~Mayer, and L.~Benini, ``Micro kinetic energy
  harvesting for autonomous wearable devices,'' in \emph{2018 International
  Symposium on Power Electronics, Electrical Drives, Automation and Motion
  (SPEEDAM)}.\hskip 1em plus 0.5em minus 0.4em\relax IEEE, 2018, pp. 105--110.

\bibitem{fan2018capturing}
K.~Fan and Z.~Liu, ``Capturing energy through a shoe-mounted piezoelectric
  energy harvester,'' in \emph{2018 IEEE/ASME International Conference on
  Advanced Intelligent Mechatronics (AIM)}.\hskip 1em plus 0.5em minus
  0.4em\relax IEEE, 2018, pp. 768--773.

\bibitem{xu2018keh}
W.~Xu, G.~Lan, Q.~Lin, S.~Khalifa, M.~Hassan, N.~Bergmann, and W.~Hu,
  ``Keh-gait: Using kinetic energy harvesting for gait-based user
  authentication systems,'' \emph{IEEE Transactions on Mobile Computing},
  vol.~18, no.~1, pp. 139--152, 2018.

\bibitem{lin2019h2b}
Q.~Lin, W.~Xu, J.~Liu, A.~Khamis, W.~Hu, M.~Hassan, and A.~Seneviratne, ``H2b:
  heartbeat-based secret key generation using piezo vibration sensors,'' in
  \emph{Proceedings of the 18th International Conference on Information
  Processing in Sensor Networks}.\hskip 1em plus 0.5em minus 0.4em\relax ACM,
  2019, pp. 265--276.

\end{thebibliography}
\end{document}